%% file: sample-authordraft.tex
\PassOptionsToPackage{table,xcdraw}{xcolor}
\documentclass[sigconf, screen, nonacm]{acmart}

\makeatletter
\def\@ACM@checkaffil{
    \if@ACM@instpresent\else
    \ClassWarningNoLine{\@classname}{No institution present for an affiliation}%
    \fi
    \if@ACM@citypresent\else
    \ClassWarningNoLine{\@classname}{No city present for an affiliation}%
    \fi
    \if@ACM@countrypresent\else
        \ClassWarningNoLine{\@classname}{No country present for an affiliation}%
    \fi
}
\makeatother

\acmConference[ISSTA 2023]{ACM SIGSOFT International Symposium on Software Testing and Analysis}{17-21 July, 2023}{Seattle, USA}

\AtBeginDocument{%
  \providecommand\BibTeX{{%
    \normalfont B\kern-0.5em{\scshape i\kern-0.25em b}\kern-0.8em\TeX}}}

\usepackage[table,xcdraw]{xcolor}
\usepackage{listings}
\usepackage{booktabs}
\usepackage{multirow}
\usepackage[most]{tcolorbox}
\usepackage{caption}
\usepackage{subcaption}
\usepackage{lipsum}
\usepackage{xcolor}
\usepackage{amsmath}
\usepackage{amsfonts}
\usepackage{acronym}
\usepackage{algorithm2e}
\def\HiLi{\leavevmode\rlap{\hbox to \hsize{\color{black!20}\leaders\hrule height .7\baselineskip depth .7ex\hfill}}}

\definecolor{darkgreen}{rgb}{0, 0.44, 0.23}
\definecolor{lightgreen}{rgb}{0.25, 0.63, 0.4375}
\definecolor{darkblue}{rgb}{0.02, 0.16, 0.49}
\newcommand{\framework}{\textsc{Eunomia}}

\newcommand{\dsl}{\textsc{Aes}}
\newcommand{\scheduler}{Scheduler}
\newcommand{\engine}{engine}
\begin{document}

\title{\framework: Enabling User-specified Fine-Grained Search in
Symbolically Executing WebAssembly Binaries}

\author{Ningyu He}
\authornote{Both first two authors contributed equally to this work.}
\affiliation{%
  \institution{Peking University}
}

\author{Zhehao Zhao}
\authornotemark[1]
\affiliation{%
  \institution{Peking University}
}

\author{Jikai Wang}
\affiliation{%
  \institution{Huazhong University of Science and Technology}
}

\author{Yubin Hu}
\affiliation{%
  \institution{Beijing University of Posts and Telecommunications}
}

\author{Shengjian Guo}
\affiliation{%
  \institution{Baidu Security}
}

\author{Haoyu Wang}
\affiliation{%
  \institution{Huazhong University of Science and Technology}
}

\author{Guangtai Liang}
\affiliation{%
  \institution{Huawei Cloud Computing Technologies Co., Ltd.}
}

\author{Ding Li}
\affiliation{%
  \institution{Peking University}
}

\author{Xiangqun Chen}
\affiliation{%
  \institution{Peking University}
}

\author{Yao Guo}
\affiliation{%
  \institution{Peking University}
}

\renewcommand{\shortauthors}{Ningyu et al.}

\begin{abstract}
Although existing techniques have proposed automated approaches to alleviate the path explosion problem of symbolic execution, users still need to optimize symbolic execution by applying various searching strategies carefully. As existing approaches mainly support only coarse-grained global searching strategies, they cannot efficiently traverse through complex code structures. In this paper, we propose {\framework}, a symbolic execution technique that allows users to specify local domain knowledge to enable fine-grained search. In {\framework}, we design an expressive DSL, {\dsl}, that lets users precisely pinpoint local searching strategies to different parts of the target program. To further optimize local searching strategies, we design an interval-based algorithm that automatically isolates the context of variables for different local searching strategies, avoiding conflicts between local searching strategies for the same variable. We implement {\framework} as a symbolic execution platform targeting WebAssembly, which enables us to analyze applications written in various languages (like C and Go) but can be compiled into WebAssembly. To the best of our knowledge, {\framework} is the first symbolic execution engine that supports the full features of the WebAssembly runtime. We evaluate {\framework} with a dedicated microbenchmark suite for symbolic execution and six real-world applications. Our evaluation shows that  {\framework} accelerates bug detection in real-world applications by up to three orders of magnitude.
According to the results of a comprehensive user study, users can significantly improve the efficiency and effectiveness of symbolic execution by writing a simple and intuitive {\dsl} script.
Besides verifying six known real-world bugs,  {\framework} also detected two new zero-day bugs in a popular open-source project, Collections-C.
\end{abstract}

\maketitle

\input{sec-i-intro.tex}

\input{sec-ii-background.tex}

\input{sec-iv-dsl.tex}

\input{sec-v-scheduler.tex}

\input{sec-engine.tex}

\input{sec-evaluation.tex}

\input{sec-case.tex}

\input{sec-viiii-related.tex}

\input{sec-conclusion.tex}


\bibliographystyle{ACM-Reference-Format}
\bibliography{sample.bib}

\end{document}

%% file: sec-i-intro.tex
\section{Introduction}
\label{sec:intro}

Symbolic execution (SE)~\cite{king1976symbolic} is a widely-used technique for detecting software vulnerabilities in OS Kernel~\cite{yang2006automatically,kim2020hfl,chipounov2009selective}, browsers~\cite{brown2020sys,fu2008safeli,ko2021oblique}, IoT firmware~\cite{yao2019identifying,hernandez2017firmusb,luo2021westworld}, blockchains~\cite{wang2020wana,singh2020blockchain,mossberg2019manticore,so2021smartest}, and other software systems~\cite{avgerinos2014enhancing,ma2011directed,cadar2011symbolic,yang2012issta}. 
However, the efficiency and scalability of symbolic execution are limited by the well-known \textit{path explosion} problem.
To accelerate symbolic execution, researchers have proposed heuristics and machine learning models to prioritize the execution of interested program paths~\cite{sharma2020java,xiao2013characteristic,GuoKWYG15,yi2015postconditioned,yang2012memoized,yi2017eliminating,li2013steering,GuoKW16,GuoWW17,cui2013verifying,sharma2020java,he2019learning,he2021learning,ruaro2021syml}.
Meanwhile, several automated techniques are also proposed to reduce the path exploration cost of the running program~\cite{ramos2015under,engler2007under,stephens2016driller,csallner2005check,brown2020sys}.
Unfortunately, the automated approaches still lack effectiveness in handling programs with complicated control flow, e.g., nested loops and multi-path loops~\cite{baldoni2018survey}.
In practice, users often need to utilize various search strategies to guide symbolic execution for their analysis goals.

Moreover, we observe that existing guiding approaches~\cite{cui2013verifying,trabish2018chopped,christakis2016guiding,brown2020sys,permenev2020verx,dockins2016constructing,he2021eosafe,cadar2008klee,wang2017angr} are often too coarse-grained to meet certain analysis purposes. 
Existing approaches mostly support a global search strategy that is applied to the whole program.
However, the global strategy is not optimal, as different code blocks of a program have distinct features that may fit different local search strategies.
Assume a nested loop is responsible for parsing received network packets. The developer wants to check if a buffer overflow exists in the nested loop. However, the inner layer has a complex function that requires a lot of time to verify. Thus, the developer may want to prioritize other light-weighted parts in the inner loop to maximize the coverage of symbolic execution. Unfortunately, existing approaches cannot prioritize a subset of the inner loop. This results in either getting stuck in processing complex functions or generating unsound analysis results. Hence, it is necessary to allow users to provide hints for \textit{local search strategies} for different program code blocks.

Towards this end, we propose {\framework}, a novel symbolic execution framework that allows users to specify fine-grained search strategies for different parts of the target program.
For example, users can specify different prioritation strategies for different layers of a nested loop.
With the help of such local search strategies, {\framework} can be several orders of magnitude faster than approaches that only support global search strategies in finding bugs. 

There are two challenges in supporting local search strategies. 
The first challenge is to effectively specify local search strategies for different parts of the target program.
To address this challenge, we propose {\dsl}, a DSL that allows users to specify local searching strategies with only a few lines of code.
{\dsl} includes a set of parameterized operations that allow users to build customized search strategies to code blocks that depend on a specific variable. Therefore, users can precisely bind the local search strategies to variables without introducing many manual efforts. The second challenge is realizing local search strategies while avoiding potential conflicts.
For example, a variable may belong to multiple code structures (e.g., shared by different layers of a nested loop). Therefore, multiple local search strategies may apply to the same variable, leading to conflicts.
To address this challenge, we propose an \textit{interval-based path searching} algorithm that automatically isolates the context of variables into intervals, naturally avoiding conflicts.

Besides the above technical contributions, this paper implements {\framework} as a symbolic execution engine targeting full-feature WebAssembly (Wasm)~\cite{wasm} binaries. Wasm is an emerging hardware-independent language that has been widely adopted by web applications~\cite{haas2017bringing,mach,magnum}, blockchain apps~\cite{park2020efficient,wasm-crypto-miner,wasm-malicious-miner}, and serverless applications~\cite{gadepalli2020sledge}.
The existing state-of-the-art symbolic execution engine for Wasm lacks full support for Wasm Interface (WASI), causing limited application scopes. To the best of our knowledge, {\framework} is the first symbolic executor that supports the full features of Wasm binaries, which could be compiled from languages such as C/C++ and Go.

We evaluate {\framework} with a widely-used micro-benchmark suite and six real-world applications from various sources, including system utilities and well-known tools written in C and Go.
In our evaluation, we show that {\framework} can reduce the execution time of symbolic execution by one to three orders of magnitude on the micro-benchmarks. In real-world applications, with the assistance of {\dsl} scripts, {\framework} detected six known real-world bugs in less than a minute.
Moreover, {\framework} has also discovered two new bugs in a real-world C library, Collections-C~\cite{collections-c}, which the developers have confirmed.
In comparison, applying only global guiding strategies can reach neither of these two bugs within two hours.
The results of a comprehensive user study also prove the utility and expressiveness of {\dsl}. It only takes 3.1 minutes for students on average for composing an effective {\dsl} script. We recruited 12 students to identify vulnerabilities 48 times in total. With \dsl, the students can trigger the wanted vulnerability 47 times within 150 seconds. On the contrary, the students can only succeed 20 times with KLEE primitives.

We summarize our main contributions as follows:
\begin{itemize}
    \item We design and implement a new symbolic execution framework, {\framework}, whose path-searching process can be tuned by user-specified domain knowledge at a fine-grained level without any modifications to the target programs.
    \item We propose a novel DSL, {\dsl}, through which users can bind a set of local fitness functions to accelerate the analysis process. Moreover, users can also introduce pre- and post-conditions for statements or functions and even to-be-checked predicates on arbitrary locations.
    \item We propose a new path search strategy, \textit{interval-based path searching}, which can isolate symbolic states into different contexts. To this end, states can be arbitrarily pruned and reordered without affecting the consistency of final results.
    \item We thoroughly evaluate {\framework} on a widely-used symbolic execution benchmark suite and several real-world applications. Moreover, we found two new vulnerabilities in in a 2.5k star GitHub project (Collections-C), which have been acknowledged and patched by the developers.
    \item To the best of our knowledge, {\framework} is the first symbolic execution framework that supports the full features of Wasm, while it also outperforms the current state-of-the-art symbolic execution tools in efficiency. 
\end{itemize}
\textbf{Availability:}  {\framework} is available at: \url{https://github.com/HNYuuu/Eunomia-ISSTA23}.

%% file: sec-ii-background.tex
\section{Motivating Example}
\label{sec:background:motivating}

We use the code snippet in Listing~\ref{lst:motivating} as a motivating example for our approach. 
This simplified example is inspired by real-world industrial network protocols~\cite{rfc7275,modbus}. 
Particularly, Listing~\ref{lst:motivating} contains a function, \texttt{check\_sections}, which takes a section vector (\texttt{sec\_vec}), and the number of fields of each vector (\texttt{sec\_field\_cnt}).
In each section, at most five fields will be used (L4).
Specifically, the first three fields, \texttt{token}, \texttt{index}, and \texttt{checksum}, present metadata of each section. They refer to the token of the sender, the index of the corresponding section in received ones, and the checksum of the following data, respectively.
The next member \texttt{len} indicates the length of the following \texttt{data}, whose correctness is validated by the above-mentioned \texttt{checksum}.
The integers stored in \texttt{sec\_field\_cnt} indicate whether the corresponding section contains data as payload or not.
The function \texttt{check\_sections} validates the fields of all sections, with the implementation in a two-layer nested loop.
The outer loop iterates all received sections, while the inner one iterates all fields and conducts the corresponding validation.

Directly running symbolic execution on Listing~\ref{lst:motivating} cannot complete the verification within a reasonable amount of time due to path explosion. In practice, developers could provide two pieces of \textit{domain knowledge} to accelerate the analysis, shown as follows: 

\begin{itemize}
    \item[\textbf{DK1}] Prioritize the less-expensive \textit{else} branch and postpone the analysis of the expensive function \texttt{foo} while verifying the user token (L23 -- L29).
    \item[\textbf{DK2}]  To avoid getting stuck in the analysis of the complex $data$ field (L41), the symbolic execution can finish analyzing the simple fields firstly, i.e., $token$, $index$, and $checksum$. 
\end{itemize}

\begin{figure*}[t!]
\centerline{\includegraphics[width=0.9\textwidth]{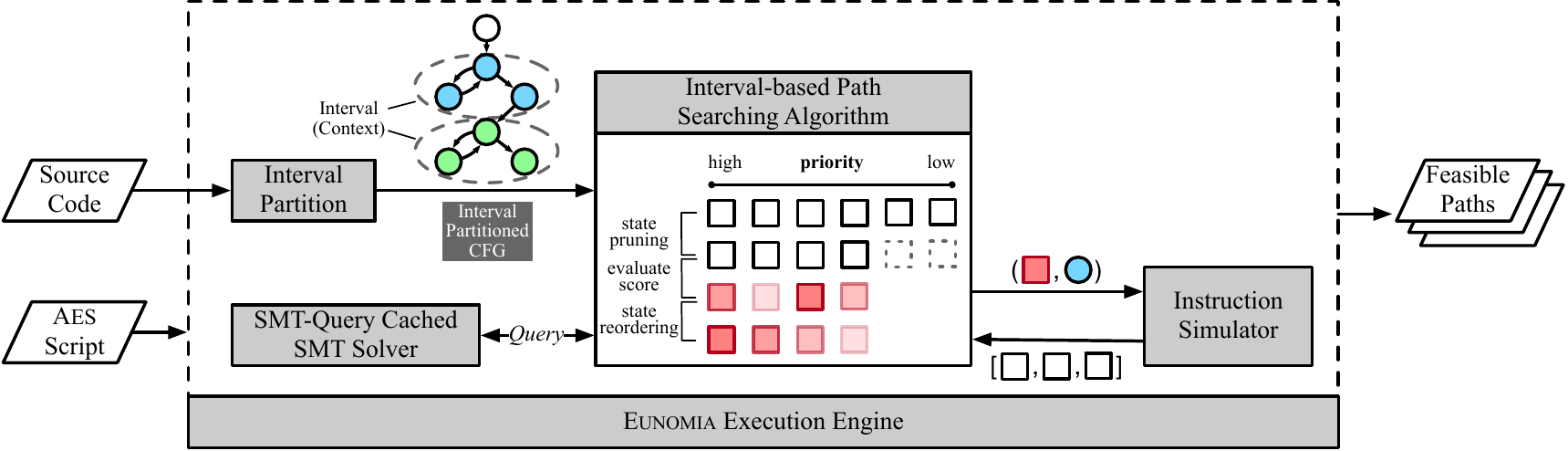}}
\caption{The workflow of {\framework}. Symbols $\square$ and {\Large$\circ$} refer to engine states and basic blocks, respectively.}
\label{fig:arch}
\end{figure*}

\begin{lstlisting}[language=C, caption={Example code for parsing received packets}, label={lst:motivating}]
#define DATA(x) *((int32_t*)(x))
#define sec_cnt 16

struct sec {
  int32_t token,
  int32_t index,
  int32_t checksum,
  int32_t len,
  char*   data
};

enum sec_name {
  TKN, IDX, CSUM, LEN, DATA
};

void check_sections(sec* sec_vec, int* sec_field_cnt) {
  int32_t i, j;
  
  for (i = 0; i < sec_cnt; i++) {
    sec* crt_sec = &(sec_vec[i]);
    // Iterate each field in section
    for (j = 0; j < sec_field_cnt[i]; j++) {
      if(j == TKN) {
        char* token = token_hash(DATA(crt_sec + j*4));
        if (isValid(token)) {
            foo();  // Expensive computation
        } else {
            bar();
        }
      } else if (j == IDX) {
        int32_t index = DATA(crt_sec + j*4);
        // Determine the correponding slot for data
      } else if (j == CSUM) {
        int32_t check_sum = DATA(crt_sec + j*4);
        assert (check_sum & 0xabcddcba) == 0x10000;
      } else if (j == LEN) {
        int32_t lenth = DATA(crt_sec + j*4);
        assert (length < 65520 && length >= 0);
        // Allocate memory for data
      } else {
        char* data = curent_sec + j*4;
        // Store data into the allocated memory
      }
    }
  }
}
\end{lstlisting}

\paragraph{\textbf{Limitations of Existing Tools}}
Unfortunately, existing tools have no effective way to utilize \textbf{DK1} and \textbf{DK2}. We take KLEE, one of the most popular symbolic execution engines as the representative to demonstrate the limitations. Other popular tools, such as CBMC~\cite{CBMC}, also suffer from similar limitations. 

KLEE cannot effectively apply \textbf{DK1} and \textbf{DK2} because it has no primitives for prioritization. Typically, we use KLEE primitives, like \texttt{klee\_assume(cond)} and \texttt{klee\_prefer\_cex(obj, cond)} for specifying extra constraints in symbolic execution. Unfortunately, those primitives can only prune unwanted states instead of prioritizing interesting paths.  
Specifically, \texttt{klee\_assume(cond)} can be used to insert extra constraints, and the paths that do not meet \texttt{cond} will be pruned.
As for \texttt{klee\_prefer\_cex(obj, cond)}, it adds a preference of values for symbolic parameters of the function to be tested.  It can be only used in the test driver instead of in any places in the code.

For \textbf{DK1}, the closest approximation that KLEE can make is to add \texttt{klee\_assume(isValid(token)==0)}, which prunes away the branches that contain \texttt{foo()}. However, In \textbf{DK1}, we only want to prioritize the branches that lead to \texttt{bar()}. Directly pruning away paths may undermine the soundness of the analysis.
Similarly, KLEE cannot utilize \textbf{DK2} either.
The closest approximation is to add \texttt{klee\_assume(j<3)} right after L22. However, this approach also compromises the soundness of the analysis since KLEE directly drops the analysis for \texttt{LEN} and \texttt{DATA} fields. 

In summary, existing tools like KLEE and CBMC lack a flexible mechanism 
to prioritize the execution of certain feasible paths, so they have limited capability to improve the execution performance by utilizing rich domain knowledge from the users.

Note that although there are other work that prioritize execution paths~\cite{ma2011directed,xie2009fitness,ruaro2021syml,he2019learning,he2021learning}, they cannot properly utilize user-defined domain knowledge as well. 
Existing path prioritization approaches either rely on pre-defined heuristics, black-box strategies, or even machine learning algorithms. Their goal is to accelerate  symbolic execution in general instead of adopting user-defined domain knowledge. Thus, they are mostly orthogonal to our work.

\paragraph{\textbf{Our Solution}}
\label{sec:method:dsl:eg}
In this section, we provide a sample code of {\dsl} that utilizes  \textbf{DK1} and \textbf{DK2} for Listing~\ref{lst:motivating}. We will discuss the formal definition of {\dsl} in \S\ref{sec:dsl}.
\lstdefinestyle{Go}{
language=Go,
morekeywords={puse, ouse, cuse, ause, call, halt, prior, def, cons, post}
}
\begin{lstlisting}[style=Go, caption={The {\dsl} script for the guidance in \S\ref{sec:background:motivating}}, label={lst:dsl}]
checker : func(check_sections) {
  // DK1: prioritize bar()
  call(bar) {prior = HIGHER;}
  // DK2: prioritize verifications on metadata fields
  puse(sec_field_cnt[i]) and puse(j) {
    prior = HIGHER if j < CSUM else LOWER;
  }
}
\end{lstlisting}

The 8-LOC {\dsl} script in Listing~\ref{lst:dsl} formalizes the DKs raised in \S\ref{sec:background:motivating}.
Two statements at L3 and L5 interpret the knowledge of \textbf{DK1} and \textbf{DK2}, respectively.
Specifically, each statement is composed of two parts, i.e., the \textit{localization part} and the \textit{knowledge part}. 
We can see that these two statements are wrapped in a \texttt{checker} that works for a function \texttt{check\_sections} (L1).
As for the \textbf{DK1}, the localization part indicates the knowledge will be attached to the position where the function \texttt{bar()} is invoked. And in the knowledge part, we can set this branch with a higher priority than the \textit{if} branch that calls \texttt{foo}.
To this end, {\framework} will first execute L28 in Listing~\ref{lst:motivating} rather than exploring both L26 and L28.

{\dsl} handles knowledge \textbf{DK2} at L5 and L6.
L5 has two \texttt{puse} expressions (ref. Figure~\ref{fig:syntax-dsl}) that localize 
the interested program point. In this case, \texttt{puse(sec\_field\_cnt[i])} and \texttt{puse(j)} 
refer to the location where both \texttt{sec\_field\_cnt[i]} and \texttt{j} are used as 
branch \textit{predicates} in the testing program.
As a result, the knowledge at L5 will be attached to the inner loop, i.e., L22 in Listing~\ref{lst:motivating}.
Then, under the context of L5, if \texttt{j} is less than \texttt{CSUM}, i.e., verifying the first three fields, 
we set those three branches with \texttt{HIGHER} priority.
Otherwise, e.g., for the symbolic states that verify \texttt{LEN} and \texttt{DATA}, we will set the priority as 
\texttt{LOWER}.
In other words, the enforced behavior by Listing~\ref{lst:dsl} is: (1) verify the first three metadata fields; 
(2) jump to the inner loop condition checking without verifying \textit{length} and \textit{data}; (3) move to 
the next section and repeat (1) \& (2); and (4) deal with the remaining \textit{length} and \textit{data} fields once 
all first three steps finish.

%% file: sec-iv-dsl.tex
\section{Design of {\framework}}
\label{sec:design}
The workflow of  {\framework} is presented in Fig.\ref{fig:arch}.
{\framework} takes the source code of the to-be-analyzed program and an {\dsl} script as input.
The CFG of the given program will be partitioned into \textit{intervals} (detailed in \S\ref{sec:background:interval}), where each of them can be regarded as an independent context.
Based on intervals, we propose an interval-based path searching algorithm.
The algorithm maintains a priority queue for states whose priority scores are evaluated by local fitness functions that are provided in {\dsl} scripts.
To this end, the algorithm pops out the state with the highest score and one of its following basic blocks as input for the instruction simulator.
The simulator conducts symbolic execution on the state according to instructions in the basic block and returns one or multiple states if path forking is necessary.
Note that states will be evaluated under their corresponding contexts.
Such an iteration continues until no candidate states are in the queue or the analysis is terminated.
{\framework} will finally output all satisfiable paths.

In the rest of this section, we will discuss  technical details of \dsl~and the {\framework} Execution Engine.

\subsection{Auxiliary {\framework} Script}
\label{sec:dsl}
{\framework} aims to help users provide fine-grained domain knowledge to accelerate the symbolic execution process.
To this end, it provides a DSL named Auxiliary {\framework} Script ({\dsl}), which allows users to specify local search strategies with a few lines of code.
Through {\dsl}, users can bind customized prioritization functions and extra constraints on states related to specific variables.

The critical challenge of {\dsl} is to provide an effective way for users to locate where to prioritize (or de-prioritize) during symbolic execution.
One straightforward method is to let users specify line numbers or functions that should be prioritized.
However, this method has two limitations.
First, it will be tedious when the user wants to specify multiple lines that follow the same pattern. Second, specifying line numbers is too coarse-grained.
Since symbolic execution propagates data and control flow dependencies of program states on the granularity of variables, a coarse-grained strategy cannot precisely locate a variable when a line of code has multiple variables and introduces ambiguity.

To avoid those limitations, we propose a more intuitive method, allowing users to bind local search strategies to variables based on their names and usage patterns. This idea partly refers to the classical \textit{def-use}~\cite{def-use} in data flow testing.
We will formally discuss the syntax and semantics of {\dsl}, and give a concrete example of {\dsl} script targeting the problem in Listing~\ref{lst:motivating}.

\subsubsection{Syntax \& Semantics of {\dsl}}
\label{sec:method:dsl:syntax}

\begin{figure}[t!]
\begin{alignat}{4}
\text{\textbf{Part I}}  &                	&&    &&        &&\nonumber\\
                        &\quad bop	\quad 	&&::\!&&= \quad && \texttt{and}\:|\:\texttt{or}\:|\:>\:|\:<\:|\:=\:|\:\geq\:|\:\leq\:|\:\neq												\nonumber\\
						&					&&    &&| \quad && +\:|\:-\:|\:\times\:|\:/\:|\:\%																				\nonumber\\
						&\quad uop	\quad 	&&::\!&&= \quad && \texttt{not}\:|\:\$																												\nonumber\\
                		&\quad l  	\quad 	&&::\!&&= \quad && \text{integer literal}\:|\:\text{float literal}\:|\:\text{char literal}													\nonumber\\
						&					&&    &&| \quad && \text{string literal}\:|\:\textbf{true}\:|\:\textbf{false}																\nonumber\\
						&\quad id  	\quad 	&&::\!&&= \quad && \textbf{halt}\:|\:\textbf{cons}\:|\:\textbf{prior}\:|\:\text{identifier}													\nonumber\\
						&\quad e  	\quad 	&&::\!&&= \quad && l\:|\:(e)\:|\:e_1\ bop\ e_2\:|\:uop\ e																\nonumber\\
						&					&&    &&| \quad && \text{identifier}\:|\:\text{identifier}[\text{identifier}]											\nonumber\\
                		&\quad s  	\quad 	&&::\!&&= \quad && e\:|\:s_1;s_2\:|\:id:=e																									\nonumber\\
						&					&&    &&| \quad && s_1\ \textbf{if}\ e\ \textbf{else}\ s_2\:|\:\textbf{while}\ e\ \textbf{do}\ s											\nonumber\\
\midrule
\text{\textbf{Part II}} &                	&&    &&        &&\nonumber\\
                        &\quad locT	\quad 	&&::\!&&= \quad && \texttt{luse}\:|\:\texttt{puse}\:|\:\texttt{cuse}\:|\:\texttt{ouse}\:|\:\texttt{ause}	\nonumber\\
						&					&&    &&| \quad && \texttt{def}\:|\:\texttt{func}\:|\:\texttt{call}										\nonumber\\
                        &\quad locE	\quad 	&&::\!&&= \quad && locT\ (\text{identifier})\:|\:locT\ (l)\:|\:locT\ (bop)																			\nonumber\\
						&					&&    &&| \quad && locE\ \texttt{and}\ locE\:|\:locE\ \texttt{or}\ locE\:|\:\texttt{not}\ locE												\nonumber\\
\midrule
\text{\textbf{Part III}}&                	&&    &&        &&\nonumber\\
                        &\quad var 		\quad 	&&::\!&&= \quad && id := e																									\nonumber\\
						&\quad behave	\quad 	&&::\!&&= \quad && locE\ \{s_1;\ldots;s_n\}																						\nonumber\\
						&					&&    &&| \quad && \texttt{pre}\ locE\ \{s_1;\ldots;s_n\}											\nonumber\\
						&					&&    &&| \quad && \texttt{post}\ locE\ \{s_1;\ldots;s_n\} 											\nonumber\\
						&\quad advice	\quad 	&&::\!&&= \quad && var\:|\:behave																								\nonumber\\
						&\quad pilot 	\quad 	&&::\!&&= \quad && \text{identifier}\ \{advice_1; \ldots;advice_n\}													\nonumber\\
						&					&&    &&| \quad && \text{identifier} : locE\ \{advice_1; \ldots;advice_n\}											\nonumber
\end{alignat}
\caption{The syntax of {\dsl}.}
\label{fig:syntax-dsl}
\end{figure}

Fig.~\ref{fig:syntax-dsl} gives the syntax of {\dsl}, which is divided into three parts according to their functionalities.
We will explain their semantics, respectively, in the following.

The terms in Part I are basic operators for specifying local guiding methods, like binary operator ($bop$), unitary operator ($uop$), literal ($l$), identifier ($id$), expression ($e$), and statement ($s$).
Two points should be noted.
First, the $\$$ in $uop$ is used to extract operands. For example, $\$0$ corresponds to the first operand of an operator.
Second, three identifiers are reserved, i.e., \texttt{halt}, \texttt{cons}, and \texttt{prior}, each of which should be followed by an expression.
Through these reserved identifiers, users can formalize the corresponding domain knowledge.
Specifically, if \texttt{halt} is set to \texttt{true}, it means that the user intends to terminate the whole analysis.
The expression that follows \texttt{cons} can be regarded as a predicate that the user wants to verify.
Finally, expressions after \texttt{prior} can be regarded as fitness functions, according to which the priority of states can be evaluated.
A concrete example of accelerating the symbolic execution process according to these three variables is illustrated in \S\ref{sec:method:dsl:eg}.

Part II contains the keywords that conduct localization via functionalities of variables.
Specifically, we design eight def-use relations in {\dsl}, which are listed in $locT$.
The \texttt{luse}, \texttt{puse}, \texttt{cuse}, \texttt{ouse}, and \texttt{ause} refer to the \textit{location use}, \textit{predicate use}, \textit{calculation use}, \textit{output use}, and \textit{argument use}, respectively.
Moreover, the \texttt{def}, \texttt{func}, and \texttt{call} correspond to \textit{variable definition or assignment}, \textit{function definition}, and \textit{function invocation}, respectively.
By applying $locT$ on literals, identifiers, and binary operators, users can specify locations precisely ($locE$).
Furthermore, $locE$ can be combined by logical operators to limit the scope.
For example, if a user intends to filter out all the ``+'' operators in the function \texttt{foo} and \texttt{bar} to verify if there are integer overflows, \texttt{(func(foo) or func(bar)) and call(+)} can meet his expectation.

Part III contains the keywords that combine the formalized domain knowledge (declared in Part I) and its corresponding positions (declared in Part II). The core term in Part III is $pilot$. 
An {\dsl} script consists of one or multiple $pilot$.
Each $pilot$ can be bound on a specific position by $locE$ to narrow down the adopted scope, typically a function like \texttt{func(foo)}.
Within a $pilot$, users can propose concrete $advice$.
Two kinds of $advice$ exist: defining auxiliary variables by $var$, or declaring concrete behaviors that should be performed on specific positions by $behave$.
Note that, a $behave$ could be further modified by two keywords: \texttt{pre} and \texttt{post}, which hint the engine the check process should be performed \textit{before} of \textit{after} the bound position, respectively.
For example, the $behave$: \texttt{post call(foo) \{cons = (i > 5);\}}, will additionally check if the \texttt{i} greater than 5 \textit{after} the invocation of the function \texttt{foo}.
Semantically, \texttt{pre} and \texttt{post} are only valid for those $behave$ with \texttt{cons} defined in.

%% file: sec-v-scheduler.tex
\subsection{{\framework} Execution Engine }
\label{sec:scheduler}

A critical challenge to realizing the local search strategies is isolating the context of {\dsl} variables. For example, the DK3 in Listing 2 intends to guide the execution of the inner loop, but the engine cannot effectively distinguish which loop the \texttt{prior} at L10 refers to.
To eliminate the ambiguity, we propose an \textit{interval-based} method that partitions a CFG into orthogonal sub-graphs, automatically isolating the variable context by sub-graphs.
In the rest of this section, we explain the definition of the interval, and propose an interval-based path searching algorithm, in which states can be pruned by user-added constraints or reordered by fitness functions.

\subsubsection{Interval}
\label{sec:background:interval}
Intuitively, an interval is a sub-graph of CFG that contains no more than one loop.
Formally, given a graph $G$ consisting of nodes $b_i$, we can define \textit{closed path} as $(b_1, \ldots, b_n)$ where edges of each adjacent pair of nodes $(b_i, b_{i+1})$ exist in $G$ and $b_1 = b_n$.
Once designating a node $h$ as a header, interval $I(h)$ is defined as \textit{the maximal, single entry sub-graph for which $h$ is the entry node and in which all closed paths contain $h$}~\cite{allen1970control}.

Given a CFG $G$, we can partition it into intervals with an iterative method~\cite{allen1970control}. The high level procedure of the algorithm is as follows:
\begin{enumerate}
	\item Initiate a queue $H$ for header nodes, and append the entry of $G$ into $H$ as $b_0$;
	\item Pop the leftmost element from $H$, say $h$, and build interval $I(h)$. Specifically, for any node $b$ in $G$, if all immediate predecessors of $b$ are already in the $I(h)$, the $b$ should also be appended into the $I(h)$. The construction of the $I(h)$ will terminate if no $b$ meets the condition;
	\item If some of (not all of) first appeared predecessors of a node $b$ are in an interval already, insert the $b$ to $H$;
	\item Pop the leftmost element from $H$, and repeat step 2 to 4 till all nodes are partitioned into intervals.
\end{enumerate}

\begin{figure}[t!]
\centering{\includegraphics{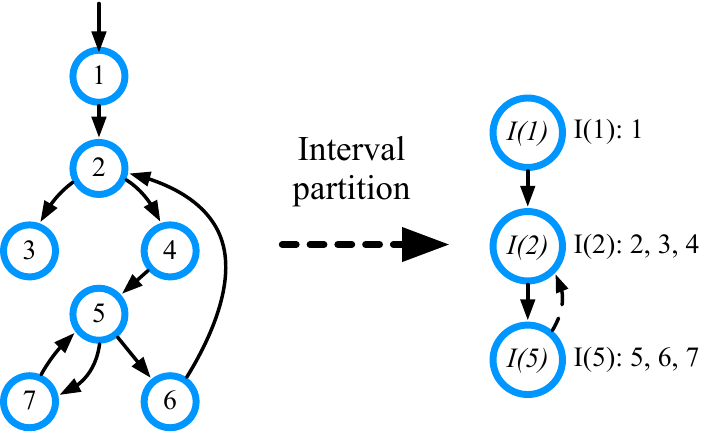}}
\caption{A CFG and its corresponding partitioned intervals.}
\label{fig:nested-loop}
\end{figure}

Take the CFG of a nested loop shown in Fig.~\ref{fig:nested-loop} as an instance to illustrate the construction process of intervals, starting from the entry, node 1.
A new interval is initiated and takes node 1 as its header, dubbed as $I(1)$.
Then, $I(1)$ tries to absorb node 2 into it.
However, since an immediate predecessor (node 6) of node 2 is not in any known intervals, a new interval should be initiated that takes it as the header ($I(2)$).
As all immediate predecessors of node 3 and node 4 are in $I(2)$ already, both of them can be included in the $I(2)$.
Similarly, node 5 will be taken as a new header, where $I(5)$ consists of nodes 5, 6, and 7.
Consequently, the CFG is partitioned by three intervals, i.e., $I(1)$, $I(2)$, and $I(5)$, whose topological relationship is sequential.
Note that, we keep the relation from node 6 to node 2, i.e., from $I(5)$ to $I(2)$, which is not an actual edge. Thus loops can be partitioned into independent intervals though for a nested loop.

\RestyleAlgo{ruled}
\begin{algorithm}[!t]
\LinesNumbered
\SetKw{Break}{break}
\SetNlSty{texttt}{}{}
\SetKwInOut{Input}{Input}\SetKwInOut{Output}{Output}
\SetKwProg{myproc}{Procedure}{}{end}
\SetKwFunction{main}{main}
\SetKwFunction{proctwo}{traverse}
\SetKwFunction{len}{len}
\SetKwFunction{append}{append}
\SetKwFunction{scorefunc}{score\_f}
\SetKwFunction{puts}{put} 
\SetKwFunction{pack}{pack}
\SetKwFunction{unpack}{unpack}
\SetKwFunction{traverse}{traverse}
\SetKwFunction{update}{update}
\SetKwFunction{succs}{succs}
\SetKwFunction{emulate}{symbolic\_execute}
\SetKwFunction{filter}{filter}
\SetKwFunction{map}{map}
\SetKwFunction{extend}{extend}
\SetKwFunction{initvar}{init\_var}
\SetKwFunction{calcheads}{calc\_interval\_heads}
\SetKwFunction{dup}{dup}
\SetKwFunction{iter}{iter}
\SetKwFunction{pop}{pop\_highest}
\SetKwFunction{clone}{clone}
\SetKwFunction{mapping}{mapping}
\SetKwFunction{queue}{queue}
\SetKwFunction{twod}{int[][]}
\SetKwFunction{stack}{stack}
\SetKwFunction{priorityqueue}{priority\_queue}
\SetKwComment{Comment}{/* }{ */}
\caption{Interval-based path searching algorithm.}\label{alg:interval-traversal}

\Input{$init\_es$: initial engine state;\\ $entry\_blk$: entry block.}
\Output{$res$: a list of possible states in accordance with all feasible paths of a function.}
\BlankLine
$heads \gets$ \calcheads{}\;
$q \gets$ \priorityqueue{}\;
\BlankLine
\myproc{\main{init\_es, entry\_blk}}{
$ms\!\gets\!(es\!=\!init\_es,cur\!=entry\_blk, pre\!=\!None,trace\!=\!\stack{},vars\!=\!\map{})$\;
$res \gets$ []\;
\HiLi$q.$\puts{\pack{priority=ms.vars\texttt{[prior]}, item=ms}}\;
\While(){$ms \gets q.$\pop{}}{
$halt,ess$ $\gets$ \traverse{ms.priority, ms.item}\;
\lIf{$halt$}{\Break}
$res.$\extend{$ess$}\;
} 
\Return{res}\;
} 
\BlankLine
\myproc(\tcc*[f]{Traverse CFG}){\proctwo{p, ms}}{
$ess\gets$\emulate{ms.es, ms.cur};\\
$succs \gets$ \succs{ms.cur}\;
\HiLi\lIf(\tcc*[f]{End of one path}){succs = $\emptyset$}{\Return{ms.vars\texttt{[halt]}, ess}} 
$T_{all} \gets ms \times ess \times succs$\;
\HiLi$T_{avail}\!\gets$ \{t~|~$t.es.cons \land ms.vars\texttt{[cons]}, \forall t\!\in\!T_{all}$\}~\;
\For{$nms, es, succ \gets$ \iter{$T_{avail}$}}{
$nms.es, nms.cur, nms.pre \gets es, succ, ms.cur$\;
\uIf{heads\texttt{[}nms.cur\texttt{]} $\neq$ heads\texttt{[}nms.pre\texttt{]}}{
    \eIf{heads\texttt{[}nms.cur\texttt{]}  $\in nms.trace$ }{
    $nms.trace,nms.vars\gets$ restore previous $trace$ and $vars$\;
    }{
        $nms.trace.$\append{heads\texttt{[}ms.cur\texttt{]}}\;
        $nms\!\gets$\!store current $vars$ and init a new one\;
    }
}
$nms.vars \gets$ update $vars$ according to {\dsl} file\;
\HiLi$q.$\puts{\pack{priority=nms.vars\texttt{[prior]}, item=nms}}\;
} 
\Return{false,\texttt{[]}}\;
} 
\end{algorithm}

\subsubsection{Interval-Based Path Searching Algorithm}
\label{sec:scheduler:traversal}
We propose a new algorithm, named \textit{interval-based path searching algorithm} (see Algorithm~\ref{alg:interval-traversal}), which can conduct symbolic execution inter- and intra-intervally.
The algorithm takes an empty engine state and the basic entry block of the given CFG as inputs. and returns states corresponding to all feasible paths as outputs.
Generally speaking, the core of the algorithm is an \textit{iteration}.
On the one side, a priority queue maintains all states with their corresponding scores calculated by the fitness functions following \texttt{prior} and dispatches the one with the highest score with its successive basic block to the {\engine}. On the other side, the {\engine} symbolically executes the given basic block, updates the state, and performs necessary forking on states which are appended into the priority queue.
Note that states may jump over different intervals, corresponding to different \dsl's contexts.
Thus context switching and restoring should be performed in the implementation of the algorithm

Delving deeper into the algorithm, the \texttt{main} function at L3 schedules all states according to their priority scores through
maintaining a data structure called \textit{meta state}, dubbed as $ms$ at L4, which packs the {\engine} returned state ($es$), current basic block and its predecessor ($cur$ and $pre$), all visited intervals' head ($trace$), and a mapping from localization expressions to the corresponding $advice$ declared in $pilot$ ($vars$).
At L6, a priority score of the initiated $ms$ will be assigned a default value, which is neither the highest nor the lowest value.
The queue will pop out the meta state with the highest priority and pass it to \texttt{traverse} to perform symbolic execution.
All feasible returned engine states will be dumped finally as outputs.

The goal of \texttt{traverse} is updating interval context and putting newly generated meta states into the priority queue.
Specifically, it first collects all forked engine states (L15), and examines if possible subsequent paths exist (L16).
If no subsequent path is obtained, indicating the current path is analyzed thoroughly, all engine states will be returned (L17).
Otherwise, to avoid unnecessary exploration, it will screen out all unsatisfiable states according to the satisfiability of collected path conditions and the predicates given by the \texttt{cons} (L19).
Then, from L21 to L29, the contexts will be updated, including visited intervals and values declared in $vars$.
Finally, a new meta state and its newly evaluated priority score will be packed and appended into the priority queue (L30).

From L21 to L30, we can sum up that the algorithm adopts a BFS-like strategy for the priority scheduling and coverage of engine states, and a DFS-like strategy for switching and restoring contexts.
If no {\dsl} script is provided, the priority queue can be considered a regular FIFO queue. Thus, the algorithm is equivalent to BFS, which guarantees the correctness of our algorithm.
Taking advantage of the characteristics of the algorithm, users can customize a fitness function in {\dsl} script to dig a loop deeper without influencing other loops' recursive time and the correctness of execution.

\subsubsection{State Scheduling}
Since the algorithm maintains states in a queue, states are independent of each other, i.e., they can be arbitrarily ordered.
Moreover, taking advantage of the characteristics of intervals, each state can possess its fitness functions (provided by the \texttt{prior} in {\dsl}) or constraints (provided by the \texttt{cons} in {\dsl}) under a non-global context.
The highlighted four lines in Algorithm~\ref{alg:interval-traversal} illustrate how states are pruned and reordered by domain knowledge provided by users.
We will detail these two processes in the following.

\vspace{0.05in}
\noindent
\textbf{Pruning Unsatisfiable States.}
The unsatisfiable engine states returned by the {\engine} will be pruned as soon as possible to improve the performance.
Except for path conditions that are collected during symbolic execution, there are also predicates provided by users through \texttt{cons}.
For instance, if a user knows the precondition of a function \texttt{bar}, that is, its argument \texttt{arg} should always be smaller than 256. The user can bind the \texttt{arg} and give a piece of advice like:
$$\texttt{pre call(bar) and ause(arg) \{cons = (arg < 256)\}}$$
State pruning will be achieved by adding the given predicates on path conditions, and verifying satisfiability by querying the backend SMT solver.

The process of state pruning is shown at L19 in Algorithm~\ref{alg:interval-traversal}.
After symbolically executing an engine state, one or several engine states are collected (L15). Also, the algorithm extracts all possible successor basic blocks as candidate blocks (L16).
Except for reaching the end of a path, all possible paths will be enumerated by a Cartesian product (L18), where $T_{all}$ can be represented as:
$$T_{all} = \{(ms, es_i, succ_j)\}, 0 \leq i \leq |ess|, 0 \leq j \leq |succs|$$
Only the tuples, whose both engine state's constraints ($t.es.cons$) and predicates defined by {\dsl} ($ms.vars[\texttt{cons}]$) are satisfiable, will be kept in $T_{avail}$.

Except for pruning unsatisfiable states, the algorithm will also consider the state assigned \texttt{halt = true} by users.
At L17, once the symbolic execution reaches the end of a path, i.e., no successive basic blocks, the engine state with its value of \texttt{halt} will be returned to the function \texttt{main}.
If the \texttt{halt} is set, the analysis will terminate immediately, which is often used in verifying the satisfiability of a property eagerly.
Note that such a customized termination on arbitrary positions does not require any modification of the running program and the framework.

\vspace{0.05in}
\noindent
\textbf{Reordering States.}
Once users provide \texttt{prior} via an {\dsl} script, the {\scheduler} can sort these states according to calculated scores each round in descending order.
In the existing symbolic execution engine, dynamic state reordering by introducing human knowledge is impractical.
Either extensive modification is required to modify the engine's path search strategy to eliminate the dependency between states, or the score of states is controlled by black-box machine learning algorithms.
As mentioned in \S\ref{sec:background:interval}, the given CFG is partitioned by independent intervals.
To this end, we can arbitrarily pick a state and run it as long as the context is changed or restored to the corresponding interval.

The implementations of state reordering locate at L6 and L30 in Algorithm~\ref{alg:interval-traversal}.
At L6, a score will be evaluated by the $advice$ bound on the current interval once a meta state is initiated.
Additionally, the loop at L20 will traverse all feasible states derived from the above pruning step.
Depending on whether the to-be-traversed interval has been accessed (L23), the context of the interval would be restored or initiated. Whichever of the two scenarios occurs, the $var$ in the interval should be updated (L29). Thus, L30 will recalculate a new score for the current interval.
Because the 2-tuple: meta state and its score, will be packed and appended into the priority queue, all the states will be reordered, and the highest one will be picked out each time at L6.

%% file: sec-engine.tex
\subsection{Implementation and Optimization}
\label{sec:engine}
We choose the WebAssembly~\cite{wasm} (Wasm) as the target language for {\framework} since it is emerging in several critical areas, including web applications~\cite{haas2017bringing,mach,magnum}, blockchain apps~\cite{park2020efficient,wasm-crypto-miner,wasm-malicious-miner}, and serverless applications~\cite{gadepalli2020sledge}.
The current state-of-the-art symbolic execution engine is a commercial open-source tool, Manticore~\cite{mossberg2019manticore}, which requires substantial manual efforts to model the APIs of Wasm runtime to analyze Wasm applications.
To ease the burden of security researchers in analyzing Wasm binaries, we implement {\framework} as the FIRST symbolic execution engine that has full support for commercial off-the-shelf Wasm applications with about 8K Python3 code.
Moreover, to ensure the efficiency of {\framework}, we propose several optimizations specified to Wasm binaries, which will be detailed in the following.

\subsubsection{Memory Modeling}
\label{sec:engine:memory}
WebAssembly adopts linear memory as the memory model.
Data in its memory is raw bit string and can be indexed and interpreted.
To emulate load and store via a concrete pointer, we adopt the mapping structure proposed by~\cite{he2021eosafe}, where the value is a raw bit string modeled by BitVector, and the key is its corresponding address range.
However, this model does not correctly deal with symbolic pointers.

To address the symbolic pointer problem~\cite{king1976symbolic}, we adopt the \textit{fully symbolic memory} model~\cite{baldoni2018survey}.
Specifically, if the loaded address is a symbol, {\framework} considers all its possible positions. Instead of forking multiple states as KLEE~\cite{cadar2008klee} does, which introduces enormous overhead, we transfer the burden to the SMT solver as it constantly updates on solving such constraints~\cite{cadar2008exe, cadar2008klee, elkarablieh2009precise, shoshitaishvili2016sok}.
In other words, we utilize \textit{if-then-else (ite)} statements to enumerate all possible positions.
For example, we need 4 bytes loaded from symbolic address $ptr_a$, where the current memory is $\{(0,5)\mapsto v\}$.
By an \texttt{ite} statement, we finally load:
$$\texttt{ite}(ptr_a = 0, v[0:4], \texttt{ite}(ptr_a = 1, v[1:5], \texttt{invalid}))$$
, where all possible addresses are iteratively taken, and the corresponding data is extracted from the BitVector $v$.
If $ptr_a$ cannot be any of the valid addresses, a symbol, \texttt{invalid}, will be returned to indicate the end of the path.
As for storing data through symbolic pointers, it works similarly. An \texttt{ite} would enumerate all feasible positions to insert the data and update the corresponding value.

\subsubsection{External Functions Emulating}
\label{sec:engine:external}
A Wasm binary is dedicated to running in a virtual environment, which plays as an intermediary between the binary and an operating system.
To this end, the {\engine} should consider the \textit{external environment problem}~\cite{baldoni2018survey}.
In the {\engine}, we apply summary-based techniques to handle this problem.
Specifically, there is a \textit{WebAssembly Interface} (WASI)~\cite{wasi}, which defines a standard interface for Wasm binaries to interact with the external environment.
WASI mainly comprises IO-related functions, like \texttt{fd\_write} and \texttt{fd\_open}.
To this end, we referred the documentation and modeled all these IO-related functions to emulate the response from the external environment.
Moreover, we also summarize behaviors of common standard library functions in C and Go, including arithmetic operations, and string and memory manipulating functions.
Consequently, all the invocations to the external will be intercepted. The corresponding fields in each state will be updated according to the function summary.

\subsubsection{SMT-Query Cache}
\label{sec:smt-cache}
Determining the satisfiability of collected constraints is a challenging problem, which is time- and resource-consuming \cite{de2008z3}.
Therefore, we have designed a cache pool for querying to alleviate this problem. Formally, we define the SMT-query cache as a set $C=[c_1, c_2,...,c_n]$ that contains all solved constraints. For each $c_i \in C$, our cache pool caches its result and all lemmas inferred from it. Then, for a given constraint $c_{solve}$ that needs to be solved, before asking SMT solvers for solving, \framework~first queries the cache $C$ following three rules:

\begin{itemize}
    \item If $c_{solve} \in C$, {\framework} directly returns the result.
    \item If $c_{solve} \not\in C \land {\exists}c_{i} \in C, c_{i} \subset c_{solve} \land c_{i}=UNSAT$, {\framework} sets $c_{solve}$ as UNSAT. 
    \item If $c_{solve} \not\in C \land {\exists}c_{i}, c_{j} \in C$, where $c_{i}, c_{j} \subset c_{solve}, |c_{i}| \geq |c_{j}|$, $c_{i}$ is the cached \textit{maximal subset} of $c_{solve}$. {\framework} first initializes the SMT solver's solving context with $c_i$ and the cached lemmas inferred from $c_i$. Then, it adds the constraint $c_{solve}-c{i}$ to the SMT solver for incremental solving and avoids calculating the results of $c_i$ again.   
\end{itemize}

If $c_{solve}$ does not match all three rules, {\framework} send $c_{solve}$ to the SMT solver and cache the result.

%% file: sec-evaluation.tex
\section{Evaluation}
\label{sec:evaluation}
We aim to evaluate the efficiency and the effectiveness of {\framework}.
Specifically, we answer the following research questions:
\begin{itemize}
\item[\textbf{RQ1}] Is {\framework} more efficient than state-of-the-art tools?
\item[\textbf{RQ2}] Is {\framework} also more effective for bug detection?
\item[\textbf{RQ3}] Is {\dsl} easy-to-use for non-expert users?
\end{itemize}

\subsection{Benchmark}
We evaluate {\framework} on both micro-benchmark programs and real-world applications.
Logic Bomb~\cite{xu2018benchmarking} is the used micro-benchmark that has 64 test cases for evaluating the performance of symbolic execution tools from 12 aspects like symbolic memory, external functions calls, floating numbers, and so on.

Our real-world application set contains six open-source applications/libraries. 
The first three are actively maintained programs that have 1.4k-16k lines of C code:
(1) Collections-C~\cite{collections-c} is a common data structures library written in C;
(2) DNSTracer~\cite{dnstracer} is a tool in Linux Kernel that determines where a Domain Name Server gets its information from for a given hostname;
(3) GOCR~\cite{gocr} is an open-sourced OCR program.
The rest three are Go projects:
(4) Snappy~\cite{snappy} is a compression tool that has 6.5K lines of code and more than 1.2K stars on Github;
(5) Go Image~\cite{go-image} is an official image manipulation library; 
and
(6)  Sprintf~\cite{go-sprintf} is the official implementation of \texttt{sprintf} function in Go.

\subsection{Experiment Setup}
\label{sec:evaluation:setup}
Our experiments are performed on a server running Ubuntu 18.04 with 16 Intel(R) Xeon(R) Platinum 8369B CPU @ 2.70GHz and 128G RAM.
We compile all targets with clang in wasi-sdk (version \texttt{14.0}) and TinyGo (version \texttt{0.21.0}).
To horizontally compare the effectiveness and efficiency brought by {\framework}, we choose Manticore~\cite{mossberg2019manticore} (version \texttt{0.3.7}) as our baseline.
Specifically, Manticore is the state-of-the-art symbolic execution engine for Wasm binaries. It is not only in commercial use and actively maintained but also open-sourced (over 3.2K stars on GitHub).
However, some additional manual efforts are necessary, or Manticore cannot directly analyze a Wasm binary.
For example, Manticore does not support imported library functions. We have to set them unreachable manually. Also, it requires an additional script in Python to emulate interactions, e.g., \texttt{getchar} and \texttt{printf}, between Wasm binaries and their external environment. Last, Manticore only regards exported functions as entries. Thus we have to export the entry for symbolic execution manually.
At last, we choose z3 (version \texttt{4.8.12}) as the back-end SMT solver because both of them support it.

\begin{table}[]
\centering
\caption{Results of symbolically executing the Logic Bomb benchmark with 5 minutes timeout limitation.}
\label{table:benchmark}
\begin{tabular}{@{}ccccc@{}}
\toprule
\textbf{}             & \begin{tabular}[c]{@{}c@{}}\#Success\\ (\%)\end{tabular} & \begin{tabular}[c]{@{}c@{}}\#Fail\\ (\%)\end{tabular}  & \begin{tabular}[c]{@{}c@{}}\#Timeout\\ (\%)\end{tabular} & \begin{tabular}[c]{@{}c@{}}\#Inapplicable\\ (\%)\end{tabular} \\ \midrule
\textbf{Manticore}    & \begin{tabular}[c]{@{}c@{}}10\\ (15.63\%)\end{tabular}   & \begin{tabular}[c]{@{}c@{}}11\\ (17.19\%)\end{tabular} & \begin{tabular}[c]{@{}c@{}}30\\ (46.88\%)\end{tabular}   & \begin{tabular}[c]{@{}c@{}}13\\ (20.31\%)\end{tabular}        \\ \midrule
\textbf{{\framework}} & \begin{tabular}[c]{@{}c@{}}29\\ (45.31\%)\end{tabular}   & \begin{tabular}[c]{@{}c@{}}9\\ (14.06\%)\end{tabular}  & \begin{tabular}[c]{@{}c@{}}13\\ (20.31\%)\end{tabular}   & \begin{tabular}[c]{@{}c@{}}13\\ (20.31\%)\end{tabular}        \\ \bottomrule
\end{tabular}
\end{table}

\begin{table}[]
\centering
\caption{Time in triggering logic bombs or bugs with 2 hours as timeout for Manticore (T$_M$), {\framework} (T$_E$), and {\framework} with the help of {\dsl} (T$_{E-dsl}$). The two bugs marked as 0-day are new bugs discovered by {\framework}.}
\label{table:effectiveness}
\resizebox{\columnwidth}{!}{%
\begin{tabular}{@{}ccccc@{}}
\toprule
                                                                               & \textbf{Vul. Type}                                                                                    & \textbf{Manticore} & \textbf{{\framework}} & \textbf{{\framework} ({\dsl})} \\ \midrule
Loop\#1                                                                        & -                                                                                                     & $\sim$101min       & 148s                  & 89s                            \\ \midrule
Loop\#2                                                                        & -                                                                                                     & $\sim$30min        & 105s                  & 66s                            \\ \midrule
Loop\#3                                                                        & -                                                                                                     & $\sim$85min        & 117s                  & 85s                            \\ \midrule
Loop\#4                                                                        & -                                                                                                     & > 2h               & > 2h                  & > 2h                           \\ \midrule
\begin{tabular}[c]{@{}c@{}}Collections-C\\ (\texttt{6f93d5})\end{tabular}      & Integer Overflow                                                                                      & > 2h               & > 2h                  & 1.5s                           \\ \midrule
\begin{tabular}[c]{@{}c@{}}Collections-C\\ (\texttt{73c468})\end{tabular}      & \begin{tabular}[c]{@{}c@{}}Implementation Error\\ in \texttt{reverse} of Array\\ (0-day)\end{tabular} & > 2h               & > 2h                  & 33s                            \\ \midrule
\begin{tabular}[c]{@{}c@{}}Collections-C\\ (\texttt{73c468})\end{tabular}      & \begin{tabular}[c]{@{}c@{}}Implementation Error\\ in \texttt{reverse} of Deque\\ (0-day)\end{tabular} & > 2h               & > 2h                  & 35s                            \\ \midrule
\begin{tabular}[c]{@{}c@{}}DNSTracer\\ (\texttt{ver.1.9})\end{tabular}         & Buffer Overflow                                                                                       & $\sim$34min        & 85s                   & 1.7s                           \\ \midrule
\begin{tabular}[c]{@{}c@{}}GOCR\\ (\texttt{ver.0.40})\end{tabular}             & Integer Overflow                                                                                      & > 2h               & $\sim$50min           & 26s                            \\ \midrule
\begin{tabular}[c]{@{}c@{}}Snappy\\ (\texttt{f4b10f})\end{tabular}             & Slice Out of Range                                                                                    & $\sim$57min        & 78s                   & 3.2s                           \\ \midrule
\begin{tabular}[c]{@{}c@{}}Go Images\\ (\texttt{72a658})\end{tabular}          & Division by Zero                                                                                      & > 2h               & $\sim$21min           & 34s                            \\ \midrule
\begin{tabular}[c]{@{}c@{}}Go \texttt{sprinf}\\ (\texttt{a2ef54})\end{tabular} & Integer Overflow                                                                                      & $\sim$25min        & 56s                   & 7s                             \\ \bottomrule
\end{tabular}%
}
\end{table}

\subsection{RQ 1: Efficiency}
\label{sec:evaluation:rq1}
To answer this question, we first compare the execution time of Manticore and {\framework} on all 64 test cases in the Logic Bomb benchmark and real-world applications with BFS global searching strategy.
Note that, because all these targets are compiled from standard toolchains, we made a few changes to adapt Manticore, as mentioned in \S\ref{sec:evaluation:setup}.
For the feasible paths analyzed by both tools on each program, we compared the number and content of the paths. This was done for two reasons: first, to confirm the correctness of the interval-based path-searching algorithm by cross-comparison; and second, to ensure that our modifications for adaptation purposes did not change the semantics of the original programs.

The results of symbolically executing the Logic Bomb benchmark are shown in Table~\ref{table:benchmark}.
As we can see, {\framework} significantly outperforms Manticore in the number of successfully triggering bomb and timeout cases.
Among all 12 categories, {\framework} has a better performance in logic bombs focusing on symbolic memory, floating numbers, and external library functions.
Though there are 13 timeout cases, we find that more than half of them are due to the lack of support of the file system. Currently, {\framework} can only step into those functions that read and write files, which is time-consuming.
Further, the inapplicable cases are mainly due to the language features of WebAssembly. Particularly, these cases include multi-threading (6), goto statement (3), socket communication (2), and asm code (2), which are not yet supported by WebAssembly.
However, some of these features have been planned in Wasm, like multi-threading in Rust to Wasm~\cite{wasm-multi}.

As for analyzing real-world applications, the results are shown in the third and fourth columns of Table~\ref{table:effectiveness}.
We can see that among eight applications, Manticore can only finish the analysis on three ones within two hours, while the number is five for {\framework}.
Moreover, it is easy to observe that {\framework} has one to two orders of magnitude improvement in efficiency compared to Manticore.

By observing the log messages in the experiment, we believe that this improvement in efficiency can be summarized in two points.
First, {\framework} adopts the memory modeling mechanism mentioned in \S\ref{sec:engine:memory}. Once encountering symbolic pointers, {\framework} will construct the corresponding \texttt{ite} statements instead of forking states with different constraints, which is time- and resource-consuming.
Second, once a set of constraints is asked for solving, Manticore will initiate multiple z3 instances with different random seeds to see which one could search for a feasible solution first.
However, {\framework} adopts the SMT-caching mechanism as we mentioned in \S\ref{sec:smt-cache}. To evaluate the improvement introduced by SMT-caching, we rerun {\framework} by disabling SMT-caching. Our experiment shows that SMT-caching can reduce the solving time by two to three orders of magnitude. We omit the detailed result in this paper due to the page limit.

\begin{tcolorbox}[title= \textbf{RQ-1} Answer, left=2pt, right=2pt, top=2pt, bottom=2pt]
Even with the default global path searching algorithm, {\framework} outperforms the state-of-the-art symbolic executor Manticore. We believe this is because some features, like the memory modeling algorithm and SMT-caching mechanism, are introduced and implemented in {\framework}.
\end{tcolorbox}

\subsection{RQ 2: Effectiveness}
\label{sec:rq2}
We compare the effectiveness of Manticore and {\framework} by measuring
the used execution time for triggering vulnerabilities in real-world cases and four loop logic bombs\footnote{These four logic bombs are the cases under the \textit{loop} category of the Logic Bomb benchmark mentioned in Table~\ref{table:benchmark}.}, as shown at the third to the fifth columns in Table~\ref{table:effectiveness}.
As we mentioned in \S\ref{sec:evaluation:rq1}, even with the identical global path searching strategy, {\framework} is tens of times better compared to Manticore in terms of bug triggering, showing excellent effectiveness in bug detection.
Moreover, with the help of local path searching strategies, which are provided by users {\dsl} scripts, the bug detection time of {\framework} can be further improved for another one to three orders of magnitude.
Take the first loop logic bomb as an example. It is an implementation of the Collatz conjecture, which takes an integer as input and conducts the following simple arithmetic operations on the input:
\begin{equation}
  \texttt{collatz(x)} =
    \begin{cases}
      x/2 & \text{if $x$ is even}\\
      3*x + 1 & \text{if $x$ is odd}
    \end{cases}       
\end{equation}

The bomb iteratively trigger the Collatz function, and can be simplified as:
\begin{lstlisting}[language=C]
int logic_bomb(int x) {
    int loopcount = 1;
    int j = collatz(x);
    while (j != 1) {
        j = collatz(j);
        loopcount++;
    }
    if (loopcount == 25) {return BOMB;}
    else {return NORMAL;}
}
\end{lstlisting}

To trigger the bomb at L8 as soon as possible, we introduce fine-grained knowledge like \texttt{puse(j) \{prior = abs(25 - loopcount);\}}.
To this end, the paths with higher \texttt{loopcount} will be prioritized.
As a result, the prioritized paths lead to a faster and hence more effective bug detection capability than Manticore.

\begin{tcolorbox}[title= \textbf{RQ-2} Answer, left=2pt, right=2pt, top=2pt, bottom=2pt]
{\framework} offers more effective bug detection capability than Manticore, no matter with the global or local search strategies. 
When introducing users domain knowledge by providing {\dsl} script, it can prioritize or defer the analysis on designated parts of a program to identify bugs effectively.
\end{tcolorbox}

\subsection{RQ 3: Usability}
\label{sec:evaluation:usability}
To evaluate the usability of {\dsl}, we have compared {\dsl} in {\framework} with the primitives in KLEE, such as \texttt{klee\_assume} and \texttt{klee\_prefer\_cex}. We achieve this by conducting a user study.

Specifically, we invited 12 computer science graduate students that did not participate in this project to learn {\dsl} and KLEE primitives. Then, we asked the students to test four test cases as listed in Table~\ref{table:effectiveness}. For each student, we first provide tutorials about KLEE and {\dsl} to them for training. Then, we hand out mini-quizzes for KLEE and {\dsl}\footnote{The quizzes can be accessed at \url{shorturl.at/enqAI}.}, respectively, to them to ensure that they have learnt the knowledge. Once the students have passed the mini-quizzes, we let them compose KLEE test drivers and {\dsl} to trigger the bugs in the given test cases. For these test cases, we told the students where to be examined and which vulnerability it has\footnote{Note that, if users adopt {\dsl} to assist the analysis on a target under real-world scenarios, they do not need to know the type and location of vulnerabilities. Part of the {\dsl} script is to examine vulnerabilities, which can be copied directly from templates. The other part can be used to guide the control flow of symbolic execution process according to their domain knowledge. See \S\ref{sec:discussion} for more details.}. 

In our user study, we evaluate three aspects. First, we evaluate the efforts needed to learn KLEE primitives and {\dsl}. This is measured by the time for the students to pass the mini-quizzes ($T_L$). Second, we evaluate the efforts required to use KLEE primitives and {\dsl}, measured by the time spent to compose the KLEE test driver and {\dsl} ($T_C$), respectively. Third, we evaluate the effectiveness of the KLEE test drivers and {\dsl}. This is measured by the execution time of the symbolic engine to find the wanted vulnerability ($T_E$). All the results are summarized in Table~\ref{table:user-study}.

\begin{table*}[]
\centering
\caption{A user study on comparing KLEE primitives and {\dsl} in {\framework} in terms of usability and executing efficiency. The $T_L$, $T_C$, and $T_E$ refer to the corresponding time consumed on syntax (l)earning, primitives/{\dsl} scripts (c)omposing, and (e)xecuting till the final results, respectively, where the \textbf{T} refers to (t)imeout that is set as 2 hours. Moreover, the parenthesis under $T_L$ refers to how many questions are answered correctly. Also, $\mathit{S_i}$ denotes a student where 
 $i\in$[1,12].}
\label{table:user-study}
\resizebox{\textwidth}{!}{%
\begin{tabular}{@{}r|cccccc|cccccc|cccccc|cccccc@{}}
\toprule
                                                                                         & \multicolumn{3}{c}{KLEE}                                                                                                            & \multicolumn{3}{c|}{{\framework}}                                                                                                    & \multicolumn{3}{c}{KLEE}                                                                                                            & \multicolumn{3}{c|}{{\framework}}                                                                                                  & \multicolumn{3}{c}{KLEE}                                                                                                          & \multicolumn{3}{c|}{{\framework}}                                                                                                  & \multicolumn{3}{c}{KLEE}                                                                                                          & \multicolumn{3}{c}{{\framework}}                                                                                                   \\
                                                                                         & $T_L$                                                                & $T_C$                          & $T_E$                       & $T_L$                                                                & $T_C$                          & $T_E$                        & $T_L$                                                                & $T_C$                          & $T_E$                       & $T_L$                                                                & $T_C$                        & $T_E$                        & $T_L$                                                                & $T_C$                        & $T_E$                       & $T_L$                                                                & $T_C$                        & $T_E$                        & $T_L$                                                                & $T_C$                        & $T_E$                       & $T_L$                                                                & $T_C$                        & $T_E$                        \\ \midrule
                                                                                         & \multicolumn{6}{c|}{$S_1$}                                                                                                                                                                                                                                                 & \multicolumn{6}{c|}{$S_2$}                                                                                                                                                                                                                                               & \multicolumn{6}{c|}{$S_3$}                                                                                                                                                                                                                                             & \multicolumn{6}{c}{$S_4$}                                                                                                                                                                                                                                              \\ \midrule
Loop\#1                                                                                  &                                                                      & 5min                           & T                           &                                                                      & 1.5min                         & 87s                          &                                                                      & 7min                           & T                           &                                                                      & 7min                         & 86s                          &                                                                      & 4min                         & T                           &                                                                      & 2min                         & 88s                          &                                                                      & 10min                        & T                           &                                                                      & 3min                         & 90s                          \\
\cellcolor[HTML]{EFEFEF}\begin{tabular}[c]{@{}r@{}}Collections-C\\ (6f93d5)\end{tabular} &                                                                      & \cellcolor[HTML]{EFEFEF}0.5min & \cellcolor[HTML]{EFEFEF}<1s &                                                                      & \cellcolor[HTML]{EFEFEF}1.5min & \cellcolor[HTML]{EFEFEF}1s   &                                                                      & \cellcolor[HTML]{EFEFEF}1min   & \cellcolor[HTML]{EFEFEF}<1s &                                                                      & \cellcolor[HTML]{EFEFEF}3min & \cellcolor[HTML]{EFEFEF}1.4s &                                                                      & \cellcolor[HTML]{EFEFEF}1min & \cellcolor[HTML]{EFEFEF}<1s &                                                                      & \cellcolor[HTML]{EFEFEF}2min & \cellcolor[HTML]{EFEFEF}1.3s &                                                                      & \cellcolor[HTML]{EFEFEF}1min & \cellcolor[HTML]{EFEFEF}<1s &                                                                      & \cellcolor[HTML]{EFEFEF}2min & \cellcolor[HTML]{EFEFEF}1.1s \\
\begin{tabular}[c]{@{}r@{}}DNSTracer\\ (ver.1.9)\end{tabular}                            &                                                                      & 1min                           & 14min                       &                                                                      & 0.5min                         & 2s                           &                                                                      & 1min                           & 14min                       &                                                                      & 3min                         & 1.7s                         &                                                                      & 1min                         & 15min                       &                                                                      & 3min                         & 1.4s                         &                                                                      & 1min                         & 14min                       &                                                                      & 4min                         & 1.4s                         \\
\cellcolor[HTML]{EFEFEF}\begin{tabular}[c]{@{}r@{}}GOCR\\ (ver.0.40)\end{tabular}        & \multirow{-6}{*}{\begin{tabular}[c]{@{}c@{}}17min\end{tabular}} & \cellcolor[HTML]{EFEFEF}2min   & \cellcolor[HTML]{EFEFEF}T   & \multirow{-6}{*}{\begin{tabular}[c]{@{}c@{}}22min\end{tabular}} & \cellcolor[HTML]{EFEFEF}0.5min & \cellcolor[HTML]{EFEFEF}26s  & \multirow{-6}{*}{\begin{tabular}[c]{@{}c@{}}23min\end{tabular}} & \cellcolor[HTML]{EFEFEF}1min   & \cellcolor[HTML]{EFEFEF}T   & \multirow{-6}{*}{\begin{tabular}[c]{@{}c@{}}27min\end{tabular}} & \cellcolor[HTML]{EFEFEF}3min & \cellcolor[HTML]{EFEFEF}T    & \multirow{-6}{*}{\begin{tabular}[c]{@{}c@{}}23min\end{tabular}} & \cellcolor[HTML]{EFEFEF}2min & \cellcolor[HTML]{EFEFEF}T   & \multirow{-6}{*}{\begin{tabular}[c]{@{}c@{}}25min\end{tabular}} & \cellcolor[HTML]{EFEFEF}3min & \cellcolor[HTML]{EFEFEF}27s  & \multirow{-6}{*}{\begin{tabular}[c]{@{}c@{}}21min\end{tabular}} & \cellcolor[HTML]{EFEFEF}3min & \cellcolor[HTML]{EFEFEF}T   & \multirow{-6}{*}{\begin{tabular}[c]{@{}c@{}}24min\end{tabular}} & \cellcolor[HTML]{EFEFEF}5min & \cellcolor[HTML]{EFEFEF}26s  \\ \midrule
                                                                                         & \multicolumn{6}{c|}{$S_5$}                                                                                                                                                                                                                                                 & \multicolumn{6}{c|}{$S_6$}                                                                                                                                                                                                                                               & \multicolumn{6}{c|}{$S_7$}                                                                                                                                                                                                                                             & \multicolumn{6}{c}{$S_8$}                                                                                                                                                                                                                                              \\ \midrule
Loop\#1                                                                                  &                                                                      & 0.5min                         & T                           &                                                                      & 4min                           & 88s                          &                                                                      & 0.5min                         & T                           &                                                                      & 5min                         & 89s                          &                                                                      & 10min                        & T                           &                                                                      & 4min                         & 149s                         &                                                                      & 3min                         & T                           &                                                                      & 1min                         & 86s                          \\
\cellcolor[HTML]{EFEFEF}\begin{tabular}[c]{@{}r@{}}Collections-C\\ (6f93d5)\end{tabular} &                                                                      & \cellcolor[HTML]{EFEFEF}0.5min & \cellcolor[HTML]{EFEFEF}<1s &                                                                      & \cellcolor[HTML]{EFEFEF}3min   & \cellcolor[HTML]{EFEFEF}1.4s &                                                                      & \cellcolor[HTML]{EFEFEF}0.5min & \cellcolor[HTML]{EFEFEF}<1s &                                                                      & \cellcolor[HTML]{EFEFEF}2min & \cellcolor[HTML]{EFEFEF}1.6s &                                                                      & \cellcolor[HTML]{EFEFEF}1min & \cellcolor[HTML]{EFEFEF}<1s &                                                                      & \cellcolor[HTML]{EFEFEF}2min & \cellcolor[HTML]{EFEFEF}1.2s &                                                                      & \cellcolor[HTML]{EFEFEF}1min & \cellcolor[HTML]{EFEFEF}<1s &                                                                      & \cellcolor[HTML]{EFEFEF}2min & \cellcolor[HTML]{EFEFEF}1.3s \\
\begin{tabular}[c]{@{}r@{}}DNSTracer\\ (ver.1.9)\end{tabular}                            &                                                                      & 1min                           & T                           &                                                                      & 3min                           & 2s                           &                                                                      & 0.5min                         & 14min                       &                                                                      & 2min                         & 1.9s                         &                                                                      & 1min                         & 14min                       &                                                                      & 3min                         & 2.2s                         &                                                                      & 2min                         & T                           &                                                                      & 2min                         & 1.3s                         \\
\cellcolor[HTML]{EFEFEF}\begin{tabular}[c]{@{}r@{}}GOCR\\ (ver.0.40)\end{tabular}        & \multirow{-6}{*}{\begin{tabular}[c]{@{}c@{}}22min\\ \end{tabular}} & \cellcolor[HTML]{EFEFEF}1min   & \cellcolor[HTML]{EFEFEF}T   & \multirow{-6}{*}{\begin{tabular}[c]{@{}c@{}}27min\\ \end{tabular}} & \cellcolor[HTML]{EFEFEF}6min   & \cellcolor[HTML]{EFEFEF}27s  & \multirow{-6}{*}{\begin{tabular}[c]{@{}c@{}}25min\\ \end{tabular}} & \cellcolor[HTML]{EFEFEF}1min   & \cellcolor[HTML]{EFEFEF}T   & \multirow{-6}{*}{\begin{tabular}[c]{@{}c@{}}26min\\ \end{tabular}} & \cellcolor[HTML]{EFEFEF}4min & \cellcolor[HTML]{EFEFEF}28s  & \multirow{-6}{*}{\begin{tabular}[c]{@{}c@{}}23min\\ \end{tabular}} & \cellcolor[HTML]{EFEFEF}3min & \cellcolor[HTML]{EFEFEF}T   & \multirow{-6}{*}{\begin{tabular}[c]{@{}c@{}}27min\\ \end{tabular}} & \cellcolor[HTML]{EFEFEF}6min & \cellcolor[HTML]{EFEFEF}31s  & \multirow{-6}{*}{\begin{tabular}[c]{@{}c@{}}22min\\ \end{tabular}} & \cellcolor[HTML]{EFEFEF}2min & \cellcolor[HTML]{EFEFEF}T   & \multirow{-6}{*}{\begin{tabular}[c]{@{}c@{}}25min\\ \end{tabular}} & \cellcolor[HTML]{EFEFEF}3min & \cellcolor[HTML]{EFEFEF}28s  \\ \midrule
\multicolumn{1}{l|}{}                                                                    & \multicolumn{6}{c|}{$S_9$}                                                                                                                                                                                                                                                 & \multicolumn{6}{c|}{$S_{10}$}                                                                                                                                                                                                                                            & \multicolumn{6}{c|}{$S_{11}$}                                                                                                                                                                                                                                          & \multicolumn{6}{c}{$S_{12}$}                                                                                                                                                                                                                                           \\ \midrule
Loop\#1                                                                                  &                                                                      & 10min                          & T                           &                                                                      & 5min                           & 86s                          &                                                                      & 2min                           & T                           &                                                                      & 5min                         & 86s                          &                                                                      & 0.5min                       & T                           &                                                                      & 5min                         & 87s                          &                                                                      & 3min                         & T                           &                                                                      & 3min                         & 90s                          \\
\cellcolor[HTML]{EFEFEF}\begin{tabular}[c]{@{}r@{}}Collections-C\\ (6f93d5)\end{tabular} &                                                                      & \cellcolor[HTML]{EFEFEF}2min   & \cellcolor[HTML]{EFEFEF}<1s &                                                                      & \cellcolor[HTML]{EFEFEF}3min   & \cellcolor[HTML]{EFEFEF}2s   &                                                                      & \cellcolor[HTML]{EFEFEF}1min   & \cellcolor[HTML]{EFEFEF}<1s &                                                                      & \cellcolor[HTML]{EFEFEF}2min & \cellcolor[HTML]{EFEFEF}1.3s &                                                                      & \cellcolor[HTML]{EFEFEF}1min & \cellcolor[HTML]{EFEFEF}<1s &                                                                      & \cellcolor[HTML]{EFEFEF}2min & \cellcolor[HTML]{EFEFEF}1.2s &                                                                      & \cellcolor[HTML]{EFEFEF}1min & \cellcolor[HTML]{EFEFEF}<1s &                                                                      & \cellcolor[HTML]{EFEFEF}2min & \cellcolor[HTML]{EFEFEF}2s   \\
\begin{tabular}[c]{@{}r@{}}DNSTracer\\ (ver.1.9)\end{tabular}                            &                                                                      & 3min                           & T                           &                                                                      & 4min                           & 1.8s                         &                                                                      & 1min                           & 15min                       &                                                                      & 4min                         & 1.3s                         &                                                                      & 0.5min                       & T                           &                                                                      & 2min                         & 1.8s                         &                                                                      & 1min                         & 15min                       &                                                                      & 2min                         & 2s                           \\
\cellcolor[HTML]{EFEFEF}\begin{tabular}[c]{@{}r@{}}GOCR\\ (ver.0.40)\end{tabular}        & \multirow{-6}{*}{\begin{tabular}[c]{@{}c@{}}20min\\ \end{tabular}} & \cellcolor[HTML]{EFEFEF}4min   & \cellcolor[HTML]{EFEFEF}T   & \multirow{-6}{*}{\begin{tabular}[c]{@{}c@{}}23min\\ \end{tabular}} & \cellcolor[HTML]{EFEFEF}5min   & \cellcolor[HTML]{EFEFEF}27s  & \multirow{-6}{*}{\begin{tabular}[c]{@{}c@{}}25min\\ \end{tabular}} & \cellcolor[HTML]{EFEFEF}3min   & \cellcolor[HTML]{EFEFEF}T   & \multirow{-6}{*}{\begin{tabular}[c]{@{}c@{}}26min\\ \end{tabular}} & \cellcolor[HTML]{EFEFEF}4min & \cellcolor[HTML]{EFEFEF}28s  & \multirow{-6}{*}{\begin{tabular}[c]{@{}c@{}}20min\\ \end{tabular}} & \cellcolor[HTML]{EFEFEF}1min & \cellcolor[HTML]{EFEFEF}T   & \multirow{-6}{*}{\begin{tabular}[c]{@{}c@{}}26min\\ \end{tabular}} & \cellcolor[HTML]{EFEFEF}3min & \cellcolor[HTML]{EFEFEF}27s  & \multirow{-6}{*}{\begin{tabular}[c]{@{}c@{}}21min\\ \end{tabular}} & \cellcolor[HTML]{EFEFEF}2min & \cellcolor[HTML]{EFEFEF}T   & \multirow{-6}{*}{\begin{tabular}[c]{@{}c@{}}25min\\ \end{tabular}} & \cellcolor[HTML]{EFEFEF}3min & \cellcolor[HTML]{EFEFEF}28s  \\ \bottomrule
\end{tabular}%
}
\end{table*}

\subsubsection{Learning and Composing Efforts}
As shown in Table~\ref{table:user-study}, the efforts required to learn {\dsl} is similar to the efforts needed to learn KLEE primitives. On average, the students used 22 minutes to pass the mini-quiz about KLEE primitives while spending 25 minutes passing the mini-quiz for {\dsl}. Although it took three minutes longer to learn {\dsl} on average, this is a one-time overhead for each user and is acceptable. 

It took 8.8 minutes for the students to create the test driver with KLEE primitives for all four cases on average. Correspondingly, the students spent 12.5 minutes to create {\dsl} scripts for all four cases on average.
Moreover, we run the Wilcoxon test over the composing time and find there is no statistical significance between KLEE and {\dsl} (p = .052). Thus, we conclude that the efforts required to compose {\dsl} are similar to the effort required to create KLEE test drivers.

\subsubsection{Expressiveness of {\dsl} and KLEE primitives.}
In general, the students can find more vulnerabilities with {\dsl} than KLEE primitives because of the powerful expressiveness of {\dsl} grammars. In Table~\ref{table:user-study}, for the 48 (12*4) test cases, the students have successfully triggered the known vulnerabilities in 47 cases within 150 seconds. The only exception is from $S_2$ where the student failed to trigger the vulnerability for GOCR. On the contrary, students can only trigger 20 vulnerabilities with KLEE primitives within the given time quota (2 hours). Specifically, none of the students can trigger the vulnerability in GOCR. This means that KLEE primitives have fundamental limitations that prohibit users from expressing effective domain knowledge.   

We further investigate why students failed to trigger the vulnerabilities in GOCR. The root reason is that KLEE cannot precisely prioritize loop branches, so it will inevitably get stuck in a vulnerability-irrelevant but computation-intensive function in GOCR. The vulnerable code of GOCR is as follows: 

\begin{lstlisting}[language=C]
int main(){
    parseImage(img, &nx, &ny);
    for(...;i < nx*ny*3;){
        // increase values of nx and ny
    }
    foo(); // computation-heavy 
}
\end{lstlisting}
Here \texttt{nx} and \texttt{ny} represent an image's width and height, respectively. This code contains an integer overflow vulnerability. In the loop body, the values of \texttt{nx} and \texttt{ny} would increase, resulting in a potential integer overflow (\texttt{nx*ny}) at L3. The effective domain knowledge for this case is to prioritize the execution of the loop body and avoid analyzing the expensive function \texttt{foo()}.

Unfortunately, neither \texttt{klee\_assume} and \texttt{klee\_prefer\_cex} can precisely express the prioritization of the loop body. Instead, according to our user study, for each of 48 cases, users can utilize the domain knowledge with at most 5 lines of {\dsl} code. We conclude that users can better utilize domain knowledge with {\dsl} than with KLEE primitives.

\begin{tcolorbox}[title= \textbf{RQ-3} Answer, left=2pt, right=2pt, top=2pt, bottom=2pt]
Compared with KLEE primitives, {\dsl} is an easy-to-learn DSL with better usability. By spending a similar amount of composing efforts, users can utilize domain knowledge and express precise prioritization with {\dsl} for faster symbolic execution. 
\end{tcolorbox}

%% file: sec-case.tex
\section{Case Study: Detecting New Bugs}
\label{sec:case}
In our experiments, {\framework} successfully identified six known bugs in real-world applications and two new extra 0-day bugs in Collections-C, as shown in the \textbf{Vul. Type} column in Table~\ref{table:effectiveness}.
In this section, we study the two new bugs in detail.

We tested all 159 interface functions provided by Collections-C by specifying their pre- and post-conditions in {\dsl}.
For each interface function, if {\framework} cannot finish the analysis in 5 minutes, we check the source code and update the {\dsl} script to determine if local search strategies can be adopted.
To better explain how {\dsl} can detect bugs, we take the bug we discovered in 
the \texttt{reverse()} function of the \texttt{deque} data structure (Listing~\ref{lst:reverse-deque}), as an instance.

\begin{lstlisting}[language=C, caption={The code of \texttt{reverse()} in deque}, label={lst:reverse-deque}]
// reverse the given deque
void reverse(CC_Deque *deque) {
    size_t i, j;
    size_t s = deque->size;
    size_t c = deque->capacity - 1;
    size_t first = deque->first;

    // the loop condition should be: i < s / 2
    for (i = 0, j = s - 1; i < (s - 1) / 2; i++, j--) {
        size_t f = (first + i) & c;
        size_t l = (first + j) & c;

        void *tmp = deque->buffer[f];
        deque->buffer[f] = deque->buffer[l];
        deque->buffer[l] = tmp;
    }
}
\end{lstlisting}

The \texttt{reverse()} function takes a deque as input and extracts the deque's size, capacity, and the first element in \texttt{s}, \texttt{c}, and \texttt{first}, respectively.
Then, in a loop, it calculates indices of the first and the last elements that are not reversed yet.
After verifying that the indices are not beyond the capacity, it will swap them and continue the loop.
However, the loop condition at L9, i.e., \texttt{i < (s - 1)}, is defective. The \texttt{i} cannot refer to the former one of the central two elements if the size of the deque is even due to the \textit{floor division}.
For symbolic execution, it is hard to trigger this bug. Because triggering this bug requires constructing a deque with symbolic length first, initiating the deque with elements, invoking \texttt{reverse()} from dozens of functions, and implementing a specific checker that examines the deque after the invocation.
Taking advantage of the expressiveness of {\dsl} and local path search strategy, we can find this bug efficiently.
Listing~\ref{lst:reverse-deque-dsl} illustrates the corresponding {\dsl} script.

\lstdefinestyle{Go}{
language=Go,
morekeywords={puse, ouse, cuse, halt, prior, def, cons, post}
}
\begin{lstlisting}[style=Go, caption={The {\dsl} script for Listing~\ref{lst:reverse-deque}}, label={lst:reverse-deque-dsl}]
checker {
  head_reverse_cnt = 0;
  tail_reverse_cnt = 0;

  func(reverse) {
    // update the counter on swapped elements
    def(f) and cuse(i) {head_reverse_cnt = head_reverse_cnt + 1;}
    def(l) and cuse(j) {tail_reverse_cnt = tail_reverse_cnt + 1;}
    
    def(s) {cons = (s >= 0 and s < 65536);}
    def(c) {cons = (c == 65535);}

    // prioritize paths heading to loop body
    cuse(s) and puse(i) {prior = HIGHER if i < (s-1)/2 else LOWER;}
  }

  post func(reverse) {cons = (head_reverse_cnt + tail_reverse_cnt == s);}
}
\end{lstlisting}

An intuition to examine the correctness of \texttt{reverse()} in deque is that: the number of swapped elements should be equal to the size of the deque if the position of the head and the tail elements can be exchanged correctly.
Therefore, based on such an intuition, we compose an {\dsl} script as shown in Listing~\ref{lst:reverse-deque-dsl}.
At L2 and L3, we declare two counters to track the number of swapped elements in the head and the tail of the deque, respectively.
They will be incremented by one once the indices of to-be-swapped elements are updated, which can be bound by \texttt{def(f) and cuse(i)} and \texttt{def(l) and cuse(j)}.
After executing \texttt{reverse()}, the sum of these two counters should be equivalent to the size of the deque, which is checked by L17.

However, without the support of the local search strategy, such a check alone may lead to false positives.
That is because the loop condition of \texttt{reverse()} (L9 at Listing~\ref{lst:reverse-deque}) is unbounded (as \texttt{s} is a symbol). Once encountering this condition, path forking is performed.
The forked path will not only lead to the path explosion problem but also result in false positives as the path that jumps out of the loop has not completed the reverse process at all, i.e., \texttt{head\_reverse\_cnt + tail\_reverse\_cnt == s} cannot be guaranteed.
Therefore, we prioritize the branch heading to the loop body by \texttt{prior = HIGHER if i < (s-1)/2 else LOWER} (the usage here is consistent with the one in \S\ref{sec:background:motivating}).
To this end, only paths that head to the loop body will be executed as they have higher priority. Once the reverse is complete, the path that jumps out of the loop can be executed, and the property at L17 will be verified.
The bug in \texttt{reverse()} of the Array data structure is similar.
These two bugs exist in the latest release, and both of them are acknowledged and patched immediately by the developer.
We urge tools that adopt Collections-C as the library to pull a new release to avoid negative impacts in their production environments.

\section{Threats to Validity}
\noindent\textbf{External Validity:} To ensure the external validity of our experiments, we select benchmarks from different independent sources. We first use Logic Bomb~\cite{xu2018benchmarking}, a well-constructed third-party benchmark suite dedicated to evaluating symbolic execution approaches. Besides, we also use six real-world applications, three written in C and three written in Go. These applications are either popular open-source projects from GitHub (e.g., with more than 1,000 stars) or the official library of Go. In general, our benchmarks can represent a wide rage of applications with different types.

\noindent\textbf{Internal Validity:} To ensure internal validity, we carefully control the parameters and variables in our experiment. For example, we configured different engines as similar as possible, e.g., using z3 as the backend in all cases. In order to evaluate the effectiveness of local searching strategies, we have compared the performance of {\framework} with global searching strategies in detail.

%% file: sec-viiii-related.tex
\section{Related Work}
\label{sec:related}

Many symbolic execution approaches were proposed for bug hunting on various targets~\cite{yang2006automatically,kim2020hfl,chipounov2009selective,brown2020sys,fu2008safeli,ko2021oblique,yao2019identifying,hernandez2017firmusb,wang2020wana,singh2020blockchain,mossberg2019manticore,so2021smartest,avgerinos2014enhancing,ma2011directed,cadar2011symbolic,yang2012issta,ChenLXGZZWL20,GuoCLCWW020,GuoCYW0LCW20,GuoWW18}.
For example, Kim et al.~\cite{kim2020hfl} have proposed HFL, combining fuzzing and symbolic execution, and found 24 previously unknown vulnerabilities in Linux kernels. He \textit{et al.}~\cite{he2021eosafe} proposed a symbolic execution engine \textsc{EOSafe} targeting EOSIO smart contracts and identified 27 in-the-wild attacks. One of the key problems is the path explosion problem. To this end, different approaches have been proposed to reduce the searching spaces of paths\cite{ramos2015under,engler2007under,stephens2016driller,csallner2005check,brown2020sys,xiao2013characteristic,baldoni2018survey,chalupa2021symbiotic,cui2013verifying,yi2015postconditioned,yang2012memoized,yi2017eliminating,li2013steering,sharma2020java,he2019learning,he2021learning,ruaro2021syml}. However, current heuristics, no matter designed manually or learned by machine-learning techniques, are only effective for specific tasks and require substantial human work if migrated to other tasks.

Researchers also proposed several techniques that guide symbolic execution with human knowledge~\cite{cui2013verifying,ma2011directed,trabish2018chopped,permenev2020verx,dockins2016constructing}. For example, Cui \textit{et al.} proposed WOODPECKER, which allows users to specify a single point of interest and automatically guide the symbolic execution to the point with program slicing~\cite{cui2013verifying}. Ma \textit{et al.} proposed guided symbolic execution, which guides symbolic execution to a target with different heuristics, such as shortest path first search or call-chain-backward search~\cite{ma2011directed}. Several DSLs are designed to specify domain-specific logic for symbolic execution. VerX combines symbolic execution with a DSL that supports temporal logic~\cite{permenev2020verx}. SAW develops a DSL for verifying properties for crypto-libraries~\cite{dockins2016constructing}. Unlike these approaches, {\dsl} is focused on modeling general and high-level human knowledge that can speed up symbolic execution. These above approaches do not offer local searching strategies, which are key contributions of {\framework}.

%% file: sec-conclusion.tex
\section{Discussion}
\label{sec:discussion}

 {\framework} is designed to better utilize users' domain knowledge instead of automatically accelerating symbolic execution. Conducting comprehensive testing on a large code base always requires lots of domain knowledge of the test programs. Therefore, introducing developers' domain knowledge is beneficial to the symbolic execution process. To make the process easy to follow, we have designed {\dsl} in an intuitive way that an ordinary programmer can learn and use.
As the results of a user study in \S\ref{sec:evaluation:usability}, without a priori knowledge, a graduate student majoring in computer science can write the corresponding {\dsl} scripts in less than 15 minutes for all the real applications.
More interestingly, as shown in \S\ref{sec:case}, the {\dsl} scripts cannot only achieve efficiency improvement but also find two 0-day bugs in real-world applications.

It is worth noting that users are not necessarily required to know the exact location and the type of vulnerability in advance.
To make the writing process of {\dsl} scripts more intuitive and user-friendly, we provide templates that can be directly re-used by other users.
For example, \texttt{post call(\$+) \{cons = (\$0 > \$1 and \$0 > \$2);\}} can bind an integer overflow detector on all additional operators.
It will automatically take effect during the symbolic execution.
Moreover, users can leverage their domain knowledge to direct the symbolic execution to key parts of the program and jump over the insignificant parts.
To this end, as the results are shown in Table~\ref{table:user-study}, {\framework} can almost outperform KLEE due to its ability to prioritize or avoid certain branches.

\section{Concluding Remarks}
\label{sec:conclusion}
In this paper, we have proposed a symbolic execution framework {\framework} that supports fine-grained local searching strategies with user-specified knowledge. We implement  {\framework} as a platform for Wasm binaries, which supports applications written in multiple mainstream languages, such as C and Go.
The experimental results show that {\framework} improves the speed of discovering bugs in real applications by up to three orders of magnitude when introducing local search strategies. Besides verifying six known bugs,  {\framework} has also discovered two zero-day bugs in a popular open-source project, Collection-C.